\documentclass[conference,letterpaper]{IEEEtran}
\normalsize
\usepackage{graphicx}
\usepackage{subcaption}
\usepackage{hyperref}
\usepackage{url}
\usepackage{array}
\usepackage[utf8]{inputenc} 
\usepackage[T1]{fontenc}
\usepackage{amsmath}
\usepackage{mathtools}
\usepackage{blkarray}
\usepackage{amsfonts}
\usepackage{amssymb}
\usepackage{amsthm}
\usepackage{enumerate}
\usepackage{cite}
\usepackage{lipsum}
\usepackage{mathtools}
\usepackage{cuted}
\usepackage{calc}
\usepackage{color}
\usepackage{epsfig}
\usepackage{setspace}
\usepackage{pstricks}
\usepackage{cancel}
\usepackage{multirow}
\usepackage{mathrsfs}
\usepackage{algorithm}
\usepackage{algpseudocode}
\usepackage{multirow}
\usepackage{diagbox}
\usepackage{multicol}
\usepackage{booktabs,makecell}
\usepackage{mathrsfs}
\usepackage{enumitem}
\newtheorem{defn}{Definition}

\newtheorem{prop}{Proposition}

\newtheorem{example}{Example}
\usepackage{geometry}
\newcommand{\W}{\mathcal{W}}

\newcommand{\B}{\mathcal{B}}

\newcommand{\R}{\mathcal{R}}
\newcommand{\T}{\mathcal{T}}

\newcommand{\M}{\mathcal{M}}
\newcommand{\G}{\mathcal{G}}

\newcommand{\V}{\mathcal{V}}

\newcommand{\X}{\mathcal{X}}
\newcommand{\s}{\mathcal{S}}
\newcommand{\D}{\mathcal{D}}

\newcommand{\J}{\mathcal{J}}

\input{epsf.sty}

\begin{document}
\title{Rate-limited  Shuffling for Distributed Computing}
\author{
	\IEEEauthorblockN{Shanuja Sasi and Onur Günlü}
	\IEEEauthorblockA{ Information Theory and Security Laboratory, Linköping University, Sweden \\
		E-mail: $\{$shanuja.sasi, onur.gunlu$\}$@liu.se}
}
\maketitle
\begin{abstract}
	This paper studies the shuffling phase in a distributed computing model with rate-limited links between nodes. Each  node  is connected to all other nodes via a noiseless broadcast link with a finite capacity. For this network, the shuffling phase is described as a distributed index-coding problem to extend an outer bound for the latter to the distributed computing problem. An inner bound on the capacity region is also established by using the distributed composite-coding scheme introduced for the distributed index-coding problem. We consider some special cases of the distributed computing problem through two examples for which we prove that the inner and outer bounds agree, thereby establishing the capacity regions. We, then, generalize the special cases to any number of nodes and computation loads under certain constraints.
\end{abstract}

\section{introduction}
\label{intro}
Distributed computing (DC) models, mainly focusing on Hadoop MapReduce \cite{Mapreduce} frameworks, are commonly used by  Google, Facebook, Amazon etc.  In the MapReduce framework, a set of servers carry out computing tasks in three phases: Map, Shuffle, and Reduce. Initially, each input data block (file) is stored multiple times across the servers, and each server processes the locally stored data to generate some intermediate values (IV) in the Map phase. In the Shuffle phase, servers exchange the IVs among themselves so that the final output functions are  distributedly calculated across the servers in the Reduce phase. Designing coding theoretic techniques to reduce the communication load during the shuffling phase has been a major field of research during the past few years
\cite{LMA,YYW,SFZ,YGK,YL,DPYTH,WCJ,DZZWL,LMAFog,PLSSM,SZZG,WCJnew,JQ,DS4,BP,SGR,survey}. 

In this paper, we consider capacity-limited links between the nodes, which, to the best of our knowledge, is not considered for the shuffling phase of the MapReduce frameworks in the literature. We connect the shuffling phase of the DC problem to distributed index-coding problem  studied in \cite{DIC,CMSIC}. We extend an outer bound on the capacity region of the distributed index-coding problem to our problem. We also derive an inner bound on the capacity region by using the distributed composite-coding scheme proposed for the distributed index-coding problem in \cite{DIC}. We consider special cases of  the DC problem, for which  we prove that the inner and outer bounds meet, thus establishing the capacity regions.

{\it Notation:} The notation $[n]$ represents the set $\{1,2, \ldots , n\}$, $[a,b]$ represents the set $\{ a, a+1, \ldots, b \}$, while $[a,b)$ represents the set $\{a,a+1, \ldots , b-1\}$. 
\section{Background and Preliminaries}
\label{problem defintion}
We consider the DC models with  MapReduce framework \cite{LMA}. In this model, there are $K$ nodes indexed by $[0,K)$. The task is to compute $Q$ output functions  $\{\phi_q:  q \in [0,Q)\}$  from $N$ distinct input files $\{w_n : n \in [0,N)\}$. Each function $\phi_q$ maps all $N$ input files, where each file $w_n \in \{0,1\}^{f}$ has $f$ bits, into a stream of $b$ bits, i.e., we have 
\begin{align}
\phi_q : \Pi_{n\in[0,N)} \{0,1\}^{f} \rightarrow \{0,1\}^{b}. 
\end{align}
 Suppose for every $q \in [0,Q)$, there is a mapping function $g_{q,n} : \{0,1\}^{f} \rightarrow \{0,1\}^{t'}$ for each $n \in [0,N)$, where $g_{q,n}$ maps the input file $w_n$ into an intermediate value (IV) $v_{q,n} = g_{q,n}(w_n) \in \{0,1\}^{t'}$ of $t'$ bits. Similarly, for every $q \in [0,Q)$, assume that there is a reduce function, $h_q : \Pi_{n\in[0,N)} \{0,1\}^{t'} \rightarrow \{0,1\}^{b}$ which maps all IVs into the output function $\phi_q = h_q(v_{q,0}, \ldots , v_{q,N-1} ) \in \{0,1\}^{b}$ of $b$ bits. With that, the output function $\phi_q$, for each  $q \in [0,Q)$, can  be equivalently described as 
\begin{align}
\phi_q(w_0, \dots , w_{N-1}) &= h_q(v_{q,0}, \ldots , v_{q,N-1} ) \nonumber \\
&= h_q(g_{q,0}(w_0), \ldots , g_{q,N-1}(w_{N-1}) ).
\end{align}
The  function computation is carried out in three phases:
\begin{enumerate}
  \item {\bf Map Phase:} The $N$ files are divided into $F$ disjoint batches,
   $\B = \{B_{f}: f \in [0,F)\}$, 
each containing $\eta_1= N / F$ 
   files, i.e., $\bigcup_{f \in [0,F)} B_{f} = \{w_0,w_1,\ldots,w_{N-1}\}$. 
    Each node $k \in [0,K)$ locally stores  subset of file batches  $\mathcal{M}_k  \subseteq \B$, and computes its IVs
    \begin{align}
   \{v_{q,n} : q \in [0,Q), w_n \in B_{f}, B_{f}  \in  \M_{k}\}.
    \end{align}
	\item {\bf Shuffle Phase:}  Each node $k \in [0,K)$ is assigned to compute a subset of output functions whose indices are in $\mathcal{W}_k \subseteq [0,Q)$.  We assume that there is a symmetric assignment across the nodes, which implies $|\mathcal{W}_k| = \eta_2= Q/K $ and $|\mathcal{W}_{k_1} \cap \mathcal{W}_{k_2}| = 0$ for all $k,k_1,k_2 \in [0,K)$ and $k_1 \neq k_2$. The set of all IVs which each node $k$ does not have access to and needs to recover for computing the assigned output functions is given
	 by  
	\begin{align}
	\{v_{q,n}: q \in \mathcal{W}_k,  B_{f}\in \B \backslash \M_{k}, w_n \in B_{f}\}.
	\end{align}
	  For each $ B_{f}\in \B \backslash \M_{k}$, we concatenate the set of IVs for the output functions in $\W_k$ which needs to be computed by node $k$ and can be computed from the files in $B_{f}$ into a message sequence 
		\begin{align}
				\label{symbols_pda}
				V_{(k,f)} = (v_{q,n} : q \in \mathcal{W}_k, w_n \in B_{f}) \in \{0,1\}^{t}
			\end{align}
		where we have $t = \eta_1 \eta_2 t'$.
	The set of all messages accessible to  node $k\in [0,K)$, which is needed by some other nodes, is given by  $ \{V_{({\hat{k},\hat{f}})}: \hat{k} \in [0,K)\backslash k,   B_{\hat{f}} \in \M_k \backslash \M_{\hat{k}} \}.$ Each  node $k$ creates a bit sequence ${Y}_k$ using these message sequences and send it through a broadcast link of capacity $C_k$ to  all the other nodes. 
	\item {\bf Reduce Phase:}  
	Receiving the sequence ${\{Y_j\}}_{j \in [0,K)\backslash k}$, each  node $k \in [0,K)$ decodes all the IVs required to compute  its output functions.
\end{enumerate}
We next define the computation load for the DC problem.
\begin{defn}
  (Computation Load \cite{LMA}): Computation load $r$ is defined as the total number of files mapped across  $K$  nodes normalized by the total number of files, i.e., we have
  \begin{align}
  	\label{cl}
  r := \frac{\sum_{k \in [0,K)} \eta_1|\M_{k}|}{N} =\frac{\sum_{k \in [0,K)} |\M_{k}|}{F}.
  \end{align}   
\end{defn}

\subsection{Distributed Index-coding Problem}
In the distributed index-coding problem \cite{DIC}, there are $M$ receivers, denoted by $ [0,M)$, a set of $M$ independent messages, $\X =\{x_0,x_1,\ldots,x_{M-1}\}$, and $2^{M}-1$ senders. Each receiver $j \in [0,M)$ wants to obtain the message $x_j$ and has a subset of messages, $S_j \subseteq \X$ as side information.  
A distributed index-coding problem can be described by a directed graph (digraph) $\G$ with $M$ vertices. Each vertex $i \in [0,M)$ represents the receiver $i$ as well as the message $x_i$ requested by the receiver $i$.
There exists an arc from a vertex $i$ to another vertex $j$ if and only if the receiver $i$ has the message $x_j$  as side-information, for $i,j \in [0,M)$.

 Let $\mathbb{M}$ denote the set of all non-empty subsets of $[0,M)$. For each $J  \in \mathbb{M}$, there is a sender that contains all the messages $\{x_j : j \in J\}$ and the broadcast link connecting sender $J$ to all the receivers has a capacity of $C_J$. Each sender $J \in \mathbb{M}$ sends a sequence $Y_J$.  Assume that each message $x_j$, for $j \in [0,M)$, is independent and uniformly distributed over the set $X_j =[2^{nR_{j}}]$, where $n$ denotes the blocklength and $R_{j}$ denotes the rate of transmission. A $((2^{nR_{j}}: j \in [0,M)), (2^{nC_J}: J \in \mathbb{M}),n)$ distributed index code is defined by a set of 
\begin{itemize}
	\item $2^{M} -1$ encoders, one at each sender $J \in \mathbb{M}$, which map the messages available at the sender $J$ into an index codeword $Y_J \in[2^{nC_J}]$ sent to  the receivers, and
	\item $M$ decoders, one at each receiver $j \in [0,M)$, which map the received sequences and side information  to a message estimate $\hat{x}_j$.
\end{itemize}
Let the estimated  messages be $\hat{\X}$. 
The average probability of error is defined as $P_e^{(n)} = \Pr[\hat{\X} \neq \X]$. A rate tuple $ (R_{j}: j \in [0,M)) $ is achievable for a given link-capacity tuple $(C_J: J \in \mathbb{M})$ if there exists a $((2^{nR_{j}}: j \in [0,M)), (2^{nC_J}: J \in \mathbb{M}),n)$ distributed index code such that $P_e^{(n)} \rightarrow 0$ as $n\rightarrow \infty$. The capacity region is the closure of the set of all achievable rate tuples.

%Both inner and outer bounds on the capacity regions were proposed for this problem in \cite{DIC}, where the inner bound was achieved by a distributed version of composite coding.  
\subsection{Multi-sender Unicast Index-Coding Problem}
\label{ms}
In multi-sender unicast index-coding problem \cite{Ong,CMSIC}, there are $M$ independent messages denoted by $\X$, $K$ senders denoted by $[0,K)$, and $M$ receivers denoted by $[0,M)$. Each receiver $j \in [0,M)$ wants to obtain the message $x_j$ and has some subset of messages, $S_j \subseteq \X$ as side information. 
Each sender $k \in [0,K)$ contains a distinct subset of messages $\X$ and is connected via a broadcast link of capacity $C_k$ to all  receivers. Therefore, with $M$ messages, the maximum number of admissible senders is $K_{max} = 2^M - 1$, and thus we have $1 \leq K\leq K_{max}$. Note that in the distributed index-coding problem \cite{DIC}, we have $K = K_{max}$ but allowed link capacity $C_k = 0$, i.e., $K_{max}$ senders are all present but some are inactive. Thus, the multi-sender unicast index-coding problem and distributed index-coding problem are equivalent.

%A new coding scheme and new bounds on capacity region for this problem were introduced using cooperative composite coding in \cite{CMSIC}.
\section{Problem Definition}
Consider the shuffling phase of the DC problem. There are $K$ senders nodes $[0,K)$ and for each sender $k \in [0,K)$, the receivers are the other nodes. 
There are $K$ receiver nodes, where each receiver node $k \in [0,K)$ wants to obtain all  messages in the set $\{V_{(k,f)}: B_f \in \B \backslash \M_k\}$, i.e., each receiver node $k$ wants $ F-|\M_k|$ number of messages. The total number of messages wanted by the nodes is given by
\begin{align}
	M=\sum_{k=0}^{K-1}F-|\M_k| = F(K-r)
\end{align}
where $r$ is the computation load as in (\ref{cl}).
Thus, there are $M = F(K-r)$ messages to be shuffled in this system, i.e., we consider only the set of all messages in the set $\{V_{(k,f)}: k \in [0,K),B_f \in \B \backslash \M_k\}$. Within this set, each node $k \in [0,K)$ knows a subset of  messages a priori, denoted by $\{V_{(\hat{k},\hat{f})}: \hat{k} \in [0,K)\backslash k,   B_{\hat{f}} \in \M_k \backslash \M_{\hat{k}} \}$. We can further divide each receiver node $k \in [0,K)$ into $ F-|\M_k|$ virtual receiver nodes indexed by $(k,f)$, for each $B_f \in  \B \backslash \M_k$. Each virtual receiver node  $(k,f)$ wants a unique message $V_{(k,f)},$  and has access to all the messages in the set $\{V_{\hat{k},\hat{f}}: \hat{k} \in [0,K)\backslash k,   B_{\hat{f}} \in \M_k \backslash \M_{\hat{k}} \}$.

For any fixed map phase configuration with computation load $r$, the shuffling phase of the DC problem is equivalent to multi-sender unicast index-coding problem  consisting of:
\begin{itemize}
	\item $M$ messages. We denote the set of all message indices involved in the system by $\V$, i.e., we have 
	\begin{align}
		\V = \{(k,f): k \in [0,K),B_f \in \B \backslash \M_k\}.	
	\end{align}
	\item $K$ sender nodes indexed by $[0,K)$. Represent the message indices available at  node $k \in [0,K)$ by $\s_k$, i.e, we have 
	\begin{align}
		\s_k &= \{({\hat{k},\hat{f}}): \hat{k} \in [0,K)\backslash k,   B_{\hat{f}} \in \M_k \backslash \M_{\hat{k}} \}.
	\end{align}
	\item $M$ virtual receiver  nodes indexed by $(k,f) \in \V$. The  message requested by the virtual receiver node $(k,f) \in \V$ is $V_{(k,f)}$.  
%	\begin{align}
%		\V_k &= \{ (k,f) : f \in [F] \backslash \M_k \} 
%	\end{align}
	The message indices available at the virtual receiver node $(k,f)$ is  given by $\s_k$.
	\item $K$ broadcast links. Each node $k \in [0,K)$ sends a sequence ${Y}_k$  to all other nodes through a noiseless broadcast channel of capacity $C_k$. 
\end{itemize}
Note that each sender node contains a distinct subset of the messages $\V$. Therefore, using similar arguments as in Section \ref{ms}, we can consider this model as a distributed index-coding problem with $K_{max} =2^{K}-1$ senders, for which only $K$ of them are active. The link capacity for the rest of the senders are assumed to be zero. Note that receiver and sender nodes consist of the same set of nodes. Throughout the paper, we refer them as receiver/sender nodes to specify the functionality of the nodes.

Following the connections established between the DC and distributed index-coding problems, a  DC problem can be described by a digraph $\G$ with $M$ vertices which represent the $M$ message indices in $\V$ and $M$ virtual receiver nodes. 
 There exists an arc from some vertex $(k_1,f_1) \in \V$ to another vertex $(k_2,f_2) \in \V$ if and only if $(k_2,f_2) \in \s_{k_1}$, i.e., when the receiver node $k_1$ has the message $V_{(k_2,f_2)}$ as side-information for $k_1,k_2 \in [0,K)$ and $f_1,f_2 \in [0,F)$ such that $k_1 \neq k_2$.
\begin{defn}
	(Data Shuffling Code): Assume that each message $V_{(k,f)} \in \V$ is independent and uniformly distributed over the set $\mathbb{V}_{(k,f)} =[2^{nR_{(k,f)}}]$, where $n$ denotes the blocklength and $R_{(k,f)}$ denotes the rate of transmission. A $((2^{nR_{(k,f)}}: (k,f) \in \V), (2^{nC_k}: k \in [0,K)),n)$ data shuffling code consists of 
	\begin{itemize}
		\item an encoder mapping at each sender node $k \in [0,K)$ which maps  messages available with it to an index codeword $Y_k \in[2^{nC_k}]$, and;
		\item a decoder mapping at each virtual receiver node $(k,f) \in \V$ which maps its received codeword symbols $\{Y_j:j \in [0,K) \backslash k\}$ and its side information messages to a requested message estimate $\hat{V}_{(k,f)}$. 
	\end{itemize}
\end{defn}
Let the estimated message indices be $\hat{\V}$. The average probability of error is defined as $P_e^{(n)} = \Pr[\hat{\V} \neq \V]$. A rate tuple $ (R_{(k,f)}: (k,f) \in \V) $ is achievable,  given link-capacities $(C_k, k \in [0,K))$, if there exists a $(2^{nR_{(k,f)}}: (k,f) \in \V), (2^{nC_k}: k \in [0,K)),n)$ data shuffling code such that $P_e^{(n)} \rightarrow 0$ as $n\rightarrow \infty$. The capacity region  is the closure of the set of achievable rate tuples.

We next 
establish bounds on the capacity region of this DC problem.
\section{Outer Bound}
In this section, we present an outer bound on the capacity
region of the DC problem  by adapting the outer
bound for the distributed index-coding problem \cite{DIC,CMSIC}. 
\begin{prop}
	\label{prop1}
	For a DC problem represented by  digraph $\G$, if the rate tuple $(R_{(k,f)}: (k,f) \in \V)$ is achievable for a given link-capacity tuple $(C_k: k \in [0,K))$, it must satisfy
	\begin{align}
		\sum_{(k,f) \in S}R_{(k,f)} \leq \sum_{j \in [0,K):S \cap \s_j \neq \phi}C_j
	\end{align}
	\noindent for all $S \subseteq \V$ for which the subgraph of $\G$ induced by $S$ does not contain a directed cycle.	 \qed
\end{prop}
The outer bound follows from Proposition 1  in \cite{CMSIC}, the proof of which is provided in [Appendix A] \cite{CMSIC}. Given a digraph $\G$, the induced acyclic subgraph obtained by removing minimum number of vertices is called the maximum acyclic induced subgraph (MAIS) of $\G$. This outer bound is a generalized version of the MAIS bound proposed for the single sender index-coding problem.
\section{Inner Bound}
The inner bound is obtained by adapting the distributed composite coding technique  based on \cite{DIC}. 

At each sender node $j \in [0, K)$, a virtual encoder is assigned for every non-empty subset of message indices $\J \subseteq \s_j$. This virtual encoder operates at an associated composite coding rate denoted as $\gamma_{\J,j}$. The encoding process consists of two steps. In the first step, the virtual encoder at sender node $j$ maps messages indexed by $\J$, denoted as $(V_{(k,f)} : (k,f) \in \J)$, into a single composite index $W_{\J,j}$. This composite index is generated randomly and independently as a Bernoulli($1/2$) sequence with a length of $2^{l_j .\gamma_{\J,j}}$ bits. In the second step, sender node $j$ utilizes flat coding to encode the composite indices $(W_{\J,j} : \J \subseteq \s_j)$ into a binary sequence $Y_{j} \in \{0, 1\}^{l_{j}}$.

Decoding also occurs in two steps. Each receiver node $k \in [0, K)$ initially recovers all composite indices $(W_{\J,j} : \J \subseteq \s_j, j \in [0, K) \backslash k)$. Error-free recovery is possible if the condition
\begin{align}
\sum\limits_{\J\subseteq \s_j:\J \nsubseteq \s_{k,j}} \gamma_{\J, j} \leq C_j
\end{align}
is satisfied, where $\s_{k,j} = \s_k \cap \s_j$ represents the common side information shared between nodes $k$ and   $j$, for $j,k \in [0,K)$.

In the second decoding step, each receiver node recovers the desired messages from the composite indices. For each $B_f \in \B \backslash \M_k$, a virtual decoder $D_{(k,f)}$ is defined at the receiver node $k \in [0,K)$. The set $\D_{(k,f),j}$ contains the messages that the decoder $D_{(k,f)}$ decodes from sender node $j$ such that $(k,f) \in \D_{(k,f),j}$. 
The probability that message $V_{(k,f)}$ can be recovered correctly at rate $R_{(k,f), j}$ goes to $1$ as $l_{j} \rightarrow \infty$, if the rates of the composite messages belong to the polymatroidal rate region $\R(D_{(k,f), j}| \s_{k,j})$ defined by 
\begin{align}
	\sum\limits_{(k,f) \in \T_{j}} R_{(k,f), j} <
\sum\limits_{\substack{\J \subseteq  \D_{(k,f),j} \cup \s_{k,j}:\\ |\J \cap \T_{j}| \neq 0} } \gamma_{\J,j}
\end{align}
for all $\T_j \subseteq \D_{(k,f),j} \backslash \s_{k,j}$ \cite{DIC}. Then, the achievable rate region for sender node $j$ is given by 
\begin{align}
\R_j \in \bigcap_{(k,f) \in \s_j} \bigcup_{\substack{\D_{(k,f),j} \subseteq \s_j :\\ (k,f) \in \D_{(k,f),j}}} \R(\D_{(k,f),j} | \s_{k,j}).
\end{align}
After establishing the composite-coding achievable rate regions $\R_j$ for all the sender nodes, we obtain a combined achievable rate region by applying the following constraints
\begin{align}
R_{(k,j)} < \sum\limits_{j \in [0,K):(k,f) \in \s_j} R_{(k,f),j} && \forall (k,f) \in \V
\end{align}
and eliminating $(R_{(k,f),j}:  j \in [0,K), (k,f) \in \V)$ via Fourier-Motzkin elimination [Appendix D]\cite{NIT}. 
		\section{Capacity Regions for Special DC Models}
			In this section, we provide the capacity regions for  special cases of the DC problem. First, we illustrate the special cases through two examples. We, then, generalize this to any $K$ and $r$ such that $(K-r)$ divides $K$.
		\begin{example}
			Consider a DC model with $3$ nodes indexed by $[0,3)$. Assume that there are $6$ input files, $\{w_0,w_1,w_2,w_3,w_4,w_5\}$ and $3$ output functions, $\{\phi_0,\phi_1,\phi_2\}$ involved. Files are divided into $3$ distinct batches, i.e., we have $\B =\{B_{0},B_{1},B_{2}\}$ such that $B_{0} =\{w_0,w_1\}, B_{1} =\{w_2,w_3\},$ and $B_{2} =\{w_4,w_5\}$.
			The set of all batches assigned to each node $k \in [0,3)$ is given by $\M_k =\{B_j: j=[0,3) \backslash k\}$. 
			Let the output function index assigned to each node $k \in [0,3)$  be $\mathcal{W}_k =\{{k}\}$. 
			
			Each node $k \in [0,3)$ can compute the IVs $\{v_{q,n}: q \in [0,3),  w_n \in B_{f}, f \in [0,3)\backslash k\}$. The set of all IVs  each node $k \in [0,3)$ does not have access to and needs to recover is 
			$\{v_{q,n}: q \in \mathcal{W}_k,  w_n \in B_{k}\}$.  We concatenate this set of IVs for the output function in $\W_k$, which needs to be computed by node $k$ and can be computed from the files in $B_{k}, $ into a message sequence 
			\begin{align}
				V_{k} = (v_{q,n} : q \in \mathcal{W}_k, w_n \in B_{k}).
			\end{align}
			Hence, the shuffling phase of this problem consists of:
			\begin{itemize}
				\item A total of $3$ messages $\{V_{0},V_{1},V_{2}\}$. Let $\V$ denote the set of all message indices, i.e., $ \mathcal{V} = \{0,1,2\}$. 
				\item $3$ sender nodes denoted by $[0,3)$. Each sender node $k \in [0,3)$ has  access to the messages whose  indices  are in $\s_{k} = \V \backslash k$, i.e., we have 
				\begin{align}
					\s_0 =\{1,2\}, \nonumber\\\s_1 =\{0,2\}, \nonumber\\\s_2 =\{0,1\}. 
				\end{align}
				\item $3$  receiver nodes denoted by $k \in \V$. Each receiver  node $k \in \V$ wants the message $V_k$ and has the other two messages, i.e., the side information set is $\s_{k}$.
			\end{itemize}
			\begin{figure}
				\centering
				\includegraphics[scale=0.52]{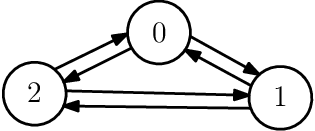}
				\caption{Digraph $\G_1$ corresponding to Example 1.}
				\label{fig: eg1}
			\end{figure}
			The digraph $\G_1$ representing this problem is shown in Fig. \ref{fig: eg1}.
			The digraph forms a clique. Hence, the MAIS for this problem contains only one vertex.
			Suppose the link capacities of the nodes are $C_0=C_1=C_2=1$. Using Proposition \ref{prop1}, an outer bound for this example is given by 
			\begin{align}
				\R_{out}^1 
				= \left. \begin{cases}
					( R_{0},R_{1},R_{2})\in \mathbb{R}_{+}^{3}:  \\ 
					\hspace{0.5cm} R_{0} \leq 2, R_{1} \leq 2, R_{2} \leq 2
				\end{cases} \right \}.
			\end{align} 
			The sender node $0$ encodes the messages $(V_{1},V_{2})$ into a composite index $W_{\{1,2\},0}$ at a rate of $\gamma_{\{1,2\},0}$. Similarly, the messages $(V_{0},V_{2})$ and $(V_{0},V_{1})$ are encoded into indices $W_{\{0,2\},1}$ and $W_{\{0,1\},2}$ at rates of $\gamma_{\{0,2\},1}$ and $\gamma_{\{0,1\},2}$ by the sender nodes $1$ and $2,$ respectively, such that $\gamma_{\{1,2\},0} \leq C_0, \gamma_{\{0,2\},1} \leq C_1,$ and $\gamma_{\{0,1\},2} \leq C_2$. The rates of the remaining indices are set to zero, i.e., $\gamma_{\{1\},0}=\gamma_{\{2\},0}=\gamma_{\{0\},1}=\gamma_{\{2\},1}=\gamma_{\{0\},2}=\gamma_{\{1\},2}=0$.

			The receiver node $0$ receives $W_{\{0,2\},1}$ and $W_{\{0,1\},2}$. Since it has side information $(V_{1},V_{2})$, it can recover $V_{0}$ from $(W_{\{0,2\},1},W_{\{0,1\},2})$ if $R_{0}< \gamma_{\{0,2\},1} + \gamma_{\{0,1\},2}$. Similarly, using  similar arguments for all  receiver nodes, a rate tuple $(R_{0},R_{1},R_{2})$ is achievable if 
			\begin{align}
				R_{0}&< \gamma_{\{0,2\},1} + \gamma_{\{0,1\},2}, \nonumber\\
				R_{1}&< \gamma_{\{0,1\},2} + \gamma_{\{1,2\},0}, \nonumber\\
				R_{2}&< \gamma_{\{0,2\},1} + \gamma_{\{1,2\},0}
			\end{align}
			for some $\gamma_{\{1,2\},0}, \gamma_{\{0,2\},1}, $ and $ \gamma_{\{0,1\},2} $ such that $\gamma_{\{1,2\},0} \leq C_0, \gamma_{\{0,2\},1} \leq C_1$ and $\gamma_{\{0,1\},2} \leq C_2$. Hence, we obtain the inequalities,
			 $R_{0} \leq 2, R_{1} \leq 2, $ and $R_{2} \leq 2$.
			Thus, the rate region $\R_{CC}^1$ achievable using the composite coding matches the outer bound $\R_{out}^1$, i.e., $\R_{CC}^1=\R_{out}^1$. \qed
		\end{example}
		\begin{example}
			Consider another DC model with $6$ nodes indexed by $[0,6)$, $6$ input files, $\{w_0,w_1,w_2,w_3,w_4,w_5\}$ and $6$ output functions, $\{\phi_0,\phi_1,\phi_2,\phi_3,\phi_4,\phi_5\}$. Divide the files into $3$ distinct batches, i.e., we have $\B =\{B_{0},B_{1},B_{2}\}$ such that $B_{0} =\{w_0,w_1\}, B_{1} =\{w_2,w_3\},$ and $B_{2} =\{w_4,w_5\}$.
			The set of all batches assigned to each node $k \in [0,6)$ is given by $\M_k =\{B_j: j\in[0,3) \backslash (k \text{ mod } 3)\}$. 
			Let the output function index assigned to each node $k \in [0,6)$  be $\mathcal{W}_k =\{{k}\}$. 
			
			Each node $k \in [0,6)$ can compute the IVs $\{v_{q,n}: q \in [0,6),  w_n \in B_{f}, f \in [0,3)\backslash (k \text{ mod } 3)\}$. 
			The set of all IVs  node $k$ does not have access and needs to recover is 
			$\{v_{q,n}: q \in \mathcal{W}_k,  w_n \in B_{(k \text{ mod } 3)} \}$.  We concatenate this set of IVs into a message sequence 
			\begin{align}
				V_{k} = (v_{q,n} : q \in \mathcal{W}_k, w_n \in B_{(k \text{ mod } 3)}).
			\end{align}
			Considering the shuffling phase, 
			\begin{itemize}
				\item There are a total of $6$ messages involved in this system $\{V_{0},V_{1},V_{2},V_{3},V_{4},V_{5}\}$. Let $\V$ denotes the set of all message indices, i.e., $ \mathcal{V} = [0,6)$. 
				\item There are $6$ sender nodes denoted by $[0,6)$ and each sender node $k \in [0,6)$ has access to the messages whose  indices  are in $\s_{k} = \V \backslash \{k, (k+3
				)	 \text{ mod } 6\}$, i.e., we have 
				\begin{align}
					\s_0 &=\{1,2,4,5\}, & \s_1 &=\{2,3,5,0\},&\s_2 &=\{3,4,0,1\}, \nonumber\\
					\s_3 &=\{4,5,1,2\}, &
					\s_4 &=\{5,0,2,3\}, & \s_5 &=\{0,1,3,4\}. 
				\end{align}
				\item There are $6$  receiver nodes denoted by $\V$. Each receiver  node $k \in \V$ wants the message $V_k$ and has all  other messages except $ V_k$ and $V_{(k+3) \text{ mod } 6}$, i.e., the side information set is $\s_{k}$.
			\end{itemize}
			\begin{figure}
				\centering
				\includegraphics[scale=0.52]{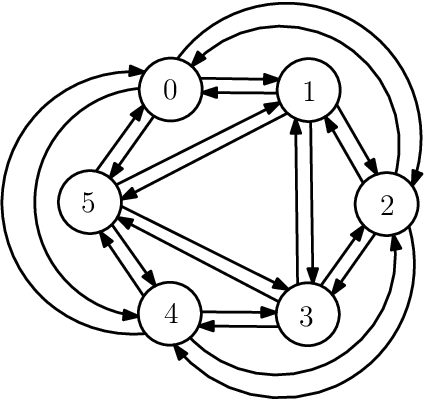}
				\caption{Digraph $\G_2$ corresponding to Example 2.}
				\label{fig: ex2}
			\end{figure}
			The digraph $\G_2$ representing this problem is shown in Fig. \ref{fig: ex2}.
			Assume the link capacities of the nodes to be $C_0=C_1=C_2=C_3=C_4=C_5=1$. 
			From  Fig. \ref{fig: ex2}, it is clear that the MAIS can contain only two vertices given by $\{k,(k+3) \text{ mod } 6\}$, for any $k \in [0,6)$.
			Using Proposition \ref{prop1}, an outer bound for this example is given by 
			\begin{align}
				\R_{out}^2
				= \left. \begin{cases}
					( R_{0},R_{1},R_{2},R_{3},R_{4},R_{5})\in \mathbb{R}_{+}^{6}:  \\ 
					\hspace{0.5cm} R_{0} + R_{3}\leq 4,R_{1} + R_{4}\leq 4, R_{2} + R_{5}\leq 4
				\end{cases} \right \}.
			\end{align} 
			The sender node $0$ encodes the messages $(V_{1},V_{2})$ into a composite index $W_{\{1,2\},0}$ at a rate of $\gamma_{\{1,2\},0}$. Similarly, the messages $V_{(k+1) \text{ mod } 6},$ and $V_{(k+2) \text{ mod } 6}$ are encoded into an index $W_{\{(k+1) \text{ mod } 6,(k+2) \text{ mod } 6\},k}$ at a rate of $\gamma_{\{(k+1) \text{ mod } 6,(k+2) \text{ mod } 6\},k}$ by the sender node $k \in [0,6)$, such that $\gamma_{\{(k+1) \text{ mod } 6,(k+2) \text{ mod } 6\},k}\leq C_k$. The rates of the remaining indices are set to  zero.
			
			The  receiver node $0$ receives $\{W_{\{5,0\},4},W_{\{0,1\},5}\}$. Since it has side information $(V_{1},V_5)$, it can recover $V_{0}$ from the composite indices if $R_{0} < \gamma_{\{5,0\},4}+\gamma_{\{0,1\},5}$. Similarly, using  similar arguments for other receiver nodes, a rate tuple $(R_{0},R_{1},R_{2},R_{3},R_{4},R_{5})$ is achievable if 
			\begin{align}
				R_{0} < \gamma_{\{5,0\},4}+\gamma_{\{0,1\},5} && R_{1} < \gamma_{\{1,2\},0}+\gamma_{\{0,1\},5} \nonumber\\
				R_{2} < \gamma_{\{1,2\},0}+\gamma_{\{2,3\},1} &&R_3< \gamma_{\{2,3\},1}+\gamma_{\{3,4\},2} \nonumber\\
				R_4< \gamma_{\{3,4\},2}+\gamma_{\{4,5\},3} &&
				R_5< \gamma_{\{4,5\},3}+\gamma_{\{5,0\},4}
			\end{align}
			for some $\gamma_{\{1,2\},0},\gamma_{\{2,3\},1},\gamma_{\{3,4\},2},\gamma_{\{4,5\},3},\gamma_{\{5,0\},4},$ and $\gamma_{\{0,1\},5}$ such that $\gamma_{\{1,2\},0}\leq C_0,\gamma_{\{2,3\},1} \leq C_1, \gamma_{\{3,4\},2} \leq C_2,\gamma_{\{4,5\},3}\leq C_3,\gamma_{\{5,0\},4}\leq C_4$ and $\gamma_{\{0,1\},5} \leq C_5$. Hence, we obtain the inequalities $R_{0} + R_{3}\leq 4,R_{1} + R_{4}\leq 4, $ and $R_{2} + R_{5}\leq 4$.
			For this example as well, the inner and outer bounds agree and we thus establish the capacity region. \qed
		\end{example}	
		We next generalize the above special cases. For the DC problem, assume that there exist some integers $K$ and $r$ such that $(K-r)$ divides $K$. Let $g=\frac{K}{K-r}$ and for any integers $a$ and $b$, let $(a)_b$ denote $(a$ mod $b)$.  In the Map phase, the input database is split into $g$ disjoint batches $\B =\{B_{k}: k \in [0,g)\}$, each containing $\eta_1 = \frac{N}{g}$ files,  i.e., $\bigcup_{k \in [0,g)} B_{k} = \{w_0,w_1,\ldots,w_{N-1}\}$. The  node $k \in [0,K)$ is assigned all batches in $\B$ except the batch $B_{(k)_g}$, i.e., $
		\M_{k} = \{B_{j} : j \in [0,g)\backslash \{(k)_g \} \},$ and
		can compute  the IVs $ \{v_{q,n} : q \in [0,Q), w_n \in B_{j},  j \in [0,g)\backslash \{(k)_g \}\}$.
		
		The set of all IVs  node $k$ does not have access to and needs to recover is given by
		$\{v_{q,n}: q \in \mathcal{W}_k,  w_n \in B_{(k)_g}\}$. 
		We concatenate the set of IVs for the output functions in $\mathcal{W}_k$ which needs to be computed by  node $k$ and can be computed from the files in $B_{(k)_g}$, i.e., $\{v_{q,n} : q \in \mathcal{W}_k, w_n \in B_{(k)_g} \}$, into the message sequence 
			\begin{align}
				V_{k} = (v_{q,n} : q \in \mathcal{W}_k, w_n \in B_{(k)_g}).
			\end{align}
			For this DC problem, the shuffling phase consists of 
			\begin{itemize}
				\item $K$ messages, whose indices are $\V =\{k: k \in [0,K) \}$.
				\item $K$ sender nodes, $[0,K)$, where each sender node $k \in [0,K)$ contains all the messages except the messages in $ \cup_{i \in [0,K-r)}V_{(k+ig)_K}$, i.e., we have
				\begin{align}
					\s_k = \left \{j: j \in [0,K) \backslash \left ( \bigcup\limits_{i \in [0,K-r)}{(k+ig)_K} \right ) \right \}.
				\end{align}
				\item $K$ receiver nodes, $[0,K)$, where each receiver node $k\in [0,K)$ requests the message $V_k$. The set of message indices available at it is given by $\s_k$.
			\end{itemize}
			This problem can be described by a digraph $\G$, with $K$ vertices which represent  $K$ message indices, and $K$ receiver nodes. Each vertex $i \in [0,K)$ represents the receiver node $i$ as well as the message $V_i$ requested by the receiver node $i$. There exists an arc from a vertex $i$ to another vertex $j$ if and only if the receiver node $i$ has the message $V_j$  as side-information for $i,j \in [0,K)$.
			
			In order to find the maximum number of vertices in the MAIS of $\G$, pick a random vertex $k$ in $\G$ first. We cannot include any vertex $j \in \{(k+u)_K:u \in [g-1]\}$, as there exists a cycle between $i$ and $j$. We pick the vertex $k+g$ next.  Continuing with similar arguments, we pick the vertices $\{(k+ig)_K:i \in [0,K-r)\}$. In short, if a vertex $i$ is included in MAIS, $g-1$ vertices before and after that vertex cannot be included in the MAIS, i.e., we cannot pick any vertices in the set $\{(k+u)_K, (k-u)_K:u \in [g-1]\}$. Hence, we can have at most $\frac{K}{g} =K-r$ vertices in the MAIS. Therefore,
			the MAIS  contains only $K-r$ vertices $\left \{ (k+ig)_K:i \in [0,K-r)\right \} $ for any $k \in [0,K)$. Using Proposition \ref{prop1}, an outer bound is given by 
			
			{\small
			\begin{align}
				\R_{out} 
				= \left. \begin{cases}
					( R_k :k\in [0,K))\in \mathbb{R}_{+}^{K}:  \\ 
					\hspace{0.5cm}\sum\limits_{i \in [0,K-r)}R_{\left (k+ig \right) _K }\leq \sum\limits_{\substack{ j \in [0,K) \backslash  \\\{(k+ig)_K: {i \in [0,K-r)}\}  } } C_j, \\ \hspace{5cm}\forall k \in [0,g)
				\end{cases} \right \}. \label{ob}
			\end{align} }
			
			\noindent For the achievability bound, each sender node $j \in [0,K)$ encodes  $g-1$ messages, whose indices are  in the set $\J = \{(j+i)_K:i\in [g-1]\}$, into a composite index $W_{\J, j}$ at a rate of $\gamma_{\J,j}$,
			such that $\gamma_{\J,j} \leq C_j$. The rates of the remaining indices are set to zero, i.e., for every $\J' \in \mathbb{S}_j \backslash \J, \gamma_{\J',j} =0$, where  $\mathbb{S}_j$ denotes the set of all non-empty subsets of $\s_j$.
			
			For each  $k \in \V$, the message index $k$ is only contained in the composite indices of the messages encoded by the sender nodes $\{(k-u)_K:u \in [g-1] \} $. 
			Hence, each receiver $k \in [0,K)$ can decode the message $V_k$ from the composite indices if
			\begin{align}
			R_k < \sum_{\substack{j \in \{(k-m_1)_K: m_1\in [g-1]\}, \\\J = \{(j+m_2)_K:\\m_2\in [g-1]\}}} \gamma_{J,j}
			\end{align}
			  which gives us the following inequalities:
			\begin{align}
			\sum\limits_{i \in [0,K-r)}R_{\left (k+ig \right)_K } &< \sum_{\substack{j \in \{(k+ig-m_1)_K: \\i \in [0,K-r),m_1\in [g-1]\}, \\\J = \{(j+m_2)_K:\\m_2\in [g-1]\}}} \gamma_{J,j}  \nonumber \\
			&\leq \sum_{\substack{j \in \{(k+ig-m_1)_K: \\i \in [0,K-r),m_1\in [g-1]\}}} C_j  \nonumber \\
			&= \sum_{\substack{j \in [0,K) \backslash \\\{(k+ig)_K:i \in [0,K-r)\}}} C_j . \label{ib}
			\end{align}
			 As (\ref{ib}) is equal to (\ref{ob}), the rate region achievable using the composite coding matches the outer bound $\R_{out}$.
\section*{Acknowledgement}
This research is supported by the ZENITH Research and Leadership Career Development Fund and the ELLIIT funding endowed by the Swedish government. We thank Parastoo Sadeghi for her insightful comments and suggestions.

\end{document}